\documentclass[reprint,amsmath,amssymb,aps]{revtex4-2}

\usepackage{graphicx}
\usepackage{amsmath,amssymb,amsthm} 
\usepackage{leftidx}
\usepackage{hyperref}
\hypersetup{colorlinks,allcolors=black}
\usepackage{floatrow}
\DeclareUnicodeCharacter{2212}{-}
\newcommand{\dd}{\mathnormal{d}}
\newcommand{\xcusp}{x_{\text{cusp}}}
\newcommand{\xc}{x_{\text{c}}}
\newcommand{\xmaxP}{x_{\text{max}}^{\text{P}}}

\newcommand{\xmaxPM}{x_{\text{max}}^{\text{P}_{\text{m}}}}
\newcommand{\betac}{\beta_{\text{c}}}
\newcommand{\xin}{x_{\text{in}}}
\newcommand{\old}[1]{}

\begin{document}

\preprint{APS/123-QED}

\title{Effect of an external mass distribution on the magnetized accretion disk}
\author{Shokoufe Faraji}
 \email{shokoufe.faraji@zarm.uni-bremen.de}
\author{Audrey Trova}%
 \email{audrey.trova@zarm.uni-bremen.de}
\affiliation{%
 University of Bremen, Center of Applied Space Technology and Microgravity (ZARM), 28359 Germany
}%
%


\begin{abstract}
In this paper, we constructed the magnetized thick disk model analytically around the static black hole in the presence of an external distribution of matter up to the quadrupole moment. This space-time is a solution to Einstein's field equation describing the exterior of a static and axially symmetric object locally. This work aims to study this space-time and the effects of quadrupole moments via studying the properties of the equilibrium sequences of magnetized, non-self-gravitating disks in this space-time. We discussed a procedure to build the thick disk model based on a combination of two approaches previously considered in the literature. We have examined different angular momentum distributions, and We have shown the properties of this relativistic accretion disc model in this background.
\end{abstract}

\maketitle


\section{Introduction}\label{intro}

Accretion disks appear in a wide variety of astrophysical phenomena, from proto-stellar disks all the way to active galactic nucleus, AGN. Accretion is supposed to be responsible for energy liberation and mass accumulation processes taking place in the core of astronomical objects.

In the accretion disk theory, one of the most significant phenomena is the role of angular momentum in constructing the models. Of course, the simplest assumption for this is to have a constant angular momentum distribution, which is the basement to introduce the thick disk model \cite{1974AcA....24...45A,1978A&A....63..209K,1980AcA....30....1J,1980A&A....88...23P,1980ApJ...242..772A,1981Natur.294..235A,1982MitAG..57...27P,1982ApJ...253..897P}. This model provides a general procedure to construct the equilibrium configurations of non-magnetized perfect fluid orbiting around a compact object. This toroidal shape disk is beneficial in the semi-analytic studies of super-Eddington accretion, also as the initial configurations in numerical simulations, e.g., \cite{1988ApJ...327..116M,1996ApJ...458..474S,1984ApJ...277..296H,1984ApJS...55..211H,2001ApJ...554L..49H,2003ApJ...592.1060D}.

There is no doubt on the role of magnetic fields in astrophysical phenomena, in particular accretion disks. This significant factor was confirmed by applying the magnetorotational instability (MRI) to accretion flows in 1991 in the work of \cite{1991ApJ...376..214B}, where the magnetic field was assumed to be dynamically weak. In this regard, \cite{Komissarov_2006} has constructed an analytically axisymmetric, stationary thick disk model with a constant distribution of specific angular momentum and a purely azimuthal magnetic field configuration. This analytical solution has some interesting astrophysical applications \cite{2015A&A...574A..48V, 2007MNRAS.378.1101M}, and is employed in the simulation set-up.
However, a more inspection in theory and simulations reveals that the accretion structure cannot be modelled by the constant specific angular moment alone. This fact led to various studies on the disk models considering different distributions of angular momentum. For instance, \cite{2009A&A...498..471Q} considered a combination of the two standard distributions, Keplerian and constant angular momentum distribution. In addition, \cite{2015MNRAS.447.3593W} extended the original set of Komissarov' solutions in the presence of the particular case of power-law distributions of angular momentum. This model is also used in studying MRI instability through time-dependent numerical simulations \cite{2017MNRAS.467.1838F}. Also, \cite{2017A&A...607A..68G} revisited the thick disk model by combining the effect of the non-constant specific angular momentum introduced in \cite{2009A&A...498..471Q} and the effect of magnetization developed in \cite{Komissarov_2006}. Another extension of the thick disk model has been developed by adding a global charge to the fluid, which interacts with a background electromagnetic field \cite{Kovar11,KoSlaCreStuKaTro16}.


These mentioned papers considered accretion onto the Kerr black hole. There are also studies on models of accretion disks that were generalized to different space-times. For example by adding a cosmological constant \cite{Stuchlik2004}, study the properties of accretion disks around naked singularities \cite{Kovacs2010}, or in wormhole space-times \cite{Harko2009}, in Schwarzschild-de Sitter black holes \cite{2003A&A...412..603R}, in Kerr-de Sitter backgrounds \cite{2005CQGra..22.3623S}, in the Reissner-Nordström-(anti-)de Sitter spacetimes \cite{2011JCAP...01..033K}, and around a boson star \cite{2021JCAP...03..063T}.

Almost all these models assumed that the space-time outside the black hole does not contain any additional matter. Nevertheless, a real astrophysical system containing a compact object is not isolated but surrounded by an accretion disk or electromagnetic fields and radiation. Since this is essential to restrict ourselves, in the first instance, to these models, the question of how an external distribution of mass may distort them is also of some interest. 

In this work, we restricted our attention to a particular class of black holes: the Schwarzschild black hole surrounded by the external axially symmetric distribution of matter due to considering quadrupole in the setup \cite{1982JMP....23..680G,Chandrasekhar:579245}. Of course, this solution is simplified to the Schwarzschild solution for vanishing external matter. However, from the point of view of the neighborhood of such a distorted object, this surrounding matter is equivalent to relaxing the assumption of asymptotic flatness. Doroshkevich et al. in 1965, first consider the Schwarzschild black hole in an external gravitational field up to a quadrupole. They constructed the metric and showed the horizon's regularity in this case \cite{1965ZhETF...49.170D}. A detailed analysis of the distorted Schwarzschild space-time's global properties was presented in the seminal paper by Geroch and Hartle \cite{1982JMP....23..680G}. They have shown that, in general, in the static, axisymmetric setup, the strong energy condition could hold, and we have an equilibrium condition for the black hole. In 2002, Chandrasekhar introduced this condition for the black hole in a static external gravitational field in terms of the multipole moments \cite{Chandrasekhar:579245}. In addition, there is a study on thin accretion disk in this space-time \cite{2020PhRvD.101b3002F}.

A survey in this background is of our interest for several reasons. First, exploring more about the exact solutions of the Einstein equation is always a valuable task. In particular, it seems that the present understanding of astronomical phenomena mostly relies on studies of stationary and axially symmetric models. Besides, it is assumed that the Schwarzschild or Kerr metrics describe astrophysical compact objects in the relativistic astrophysical study. However, besides these setups, others can imitate a black hole's properties, such as the electromagnetic signature \citep{PhysRevD.78.024040}. In addition, astrophysical observations may not be fitted, in general, within the general theory of relativity by using the Schwarzschild or Kerr metric \citep{2019MNRAS.482...52S,2002A&A...396L..31A}.

Moreover, this setup could constitute a reasonable model of a real situation that arises in this compact object's vicinity with the possibility of analytic analysis via exercising parameters of models where can be treated as the degrees of freedom of the system.  Above all, maybe the main physical motivation for analytical and semi-analytic studies of non-accreting structures seems to explain certain qualitative features of expected or actual numerical accretion disc models and this study can apply in this area.

This set up in some sense, is similar to the Schwarzschild solution with an additional external gravitational field, similar to adding a magnetic environment to the black hole solution \cite{1976JMP....17...54E,1976JMP....17..182E,PhysRevD.10.1680}. Or it can model the effect of the accretion disk's outer part on its inner part to study the self-gravity of the disk, which is a work in progress.
In this work, we study this solution by examining the thick disk model structure in this space-time. For this purpose, we generalized the extension of the models in some respects. First, combine the approaches described in \citep{Komissarov_2006}, and \citep{2009A&A...498..471Q}. Second, consider \citep{Komissarov_2006}, and \citep{2015MNRAS.447.3593W} to construct different magnetised disc models with non-constant angular momentum distributions, where in this procedure, the location and morphology of the equipotential surfaces can be computed numerically. Our main motivation to consider the non-constant angular momentum distribution was due to the fact that, in general, detailed features of the accretion structure cannot be modelled considering only either the Keplerian or the constant one. The first choices could be either utilizing power-law distribution, which is mathematically a simple and reasonable model, or a combined distribution of the Keplerian and the constant angular momentum. This simple model can lead to a wider variety of possible accretion models, which can be served in numerical simulations. As we shall show, for the particular case of constant angular momentum distributions and vanishing quadrupole moment, the results are in good agreement with the results of Komissarov. We shall also compare the result of non-constant angular momentum with the particular case of constant angular momentum distributions.


We adopted the convention $(-+++)$ for signature, also geometrical unit $c=G=1$, unless for presenting the results in Section \ref{resultsd}. The organization of the paper is as follows: Section \ref{sec1} presents the analytic framework to build the thick disks while Section \ref{sec2} explains the magnetized version and different angular momentum distributions. Section \ref{sec3} is dedicated to explaining the space-time structure briefly. The results and discussion are presented in Section \ref{resultsd}. Finally, summary and conclusion presented in Section \ref{sec:discuss}.

\section{Thick disk model}
\label{sec1}
The thick disk is the simplest analytic model of the hydrodynamical structure of an accretion disk with no accretion flow based on the Boyer's condition \footnote{Boyer's condition states the boundary of any stationary and barotropic perfect fluid body is an equipotential surface.}.

In this model, the equation of state is taken to be barotropic, and self-gravity is negligible. We briefly explain this model in the static case that we have studied. The metric, in the spherical coordinates, reads as

\begin{align}
    {\rm d}s^2=g_{tt}{\rm d}t^2+g_{rr}{\rm d}r^2+g_{\theta \theta}{\rm d}{\theta}^2+g_{\phi \phi}{\rm d}{\phi}^2.
\end{align}
where components of metric only depend on $r$ and $\theta$. Also, rotation of perfect fluid is assumed to be in the azimuthal direction; therefore the four-velocity and stress-energy tensor simplifies to
\begin{align}
    u^{\mu}&=(u^t,0,0,u^{\phi}),\\
    T^{\mu}{}_{\nu}&=wu_{\nu}u^{\mu}-\delta^{\mu}{}_{\nu}p,\
\end{align}
here $w$ is the enthalpy, and $p$ is pressure. In $ T^{\mu}{}_{\nu}$ the dissipation due to the viscosity and the heat conduction are neglected.
Adopting the definition of the specific angular momentum and the angular velocity in the static case, we have

\begin{equation}
\ell=-\frac{u_{\phi}}{u_{t}}, \quad \Omega=\frac{u^{\phi}}{u^{t}},
\end{equation}
and their relationship as

\begin{equation}\label{omel}
\ell=-\frac{\Omega g_{\phi \phi}}{g_{t t}}, \quad \Omega=-\frac{\ell g_{t t}}{g_{\phi \phi}}.
\end{equation}
Also, the corresponding redshift factor in this set up is given by

\begin{align}
(u_t)^{-2}=\frac{\ell^2 g_{tt}+g_{\phi\phi}}{g_{tt}g_{\phi\phi}}.
\end{align}
The relativistic Euler equation is then written as \cite{1978A&A....63..221A},

\begin{align}\label{maintori}
    \int_{p_{\rm in}}^{p}{\frac{{\rm d}p}{w}}=-\ln{|u_t|}+\ln{|(u_t)_{\rm in}|}+\int_{\ell_{\rm in}}^{\ell}{\frac{\Omega{\rm d}\ell}{1-\Omega\ell}},
\end{align}
where the subscript \textit{in} refers to the inner edge of the disk. This relation implies $\Omega=\Omega(\ell)$ and satisfies the general relativistic version of the von Zeipel theorem, which states for a toroidal magnetic field, the surfaces of constant $p$ coincide with constant $w$, if and only if constant $\Omega$ and constant $\ell$ coincide \cite{1924MNRAS..84..665V, 2015GReGr..47...44Z}. However, for the non magnetized version, the surface of equal $\Omega$, $\ell$, $p$ and $w$ all coincide \cite{1978A&A....63..221A}. Therefore, by specifying $\Omega=\Omega(\ell)$, one can construct the model via solving equation \eqref{omel} for $\Omega$ or $\ell$. Then from equation \eqref{maintori}, one obtains $W(r,\theta)$ and $p(r,\theta)$. This model can be adapted for either constant or a non-constant angular momentum distribution. We explore more about this while constructing the magnetized version.

\section{Magnetized thick disk model}
\label{sec2}
In the magnetized case, all previous assumptions are indefeasible. Besides, following \cite{Komissarov_2006} the magnetic field is assumed to be purely azimuthal, i.e., the four-vector magnetic field has vanishing $r$ and $\theta$ components, and the flow is axially symmetric and stationary. In this method, we employ the strong toroidal magnetic field pressure compared to the gas pressure \cite{2015MNRAS.447.3593W}. In what follows, the equations and assumptions of the magnetic thick disk are reviewed.

\subsection{ Equations of ideal relativistic MHD}
\label{sec:level11}
Conservation laws that describe the covariant equations governing the evolution of the gas in the ideal relativistic Magnetohydrodynamics, MHD, are baryon conservation, stress-energy conservation, and induction equation \cite{1978srfm.book.....D,1989rfmw.book.....A},

\begin{align}\label{eq:Property}
\left(\rho u^{\nu}\right)_{;\nu}&=0 \,,\\
T^{\nu\mu}{}_{;\nu}&=0 \,, \\
\leftidx{^*}{F}{^{\nu \mu}}{}_{;\nu} &=0 \,,
\end{align}
where subscript $";"$ refers to the covariant derivative, $T^{\mu \nu}$ is the total stress-energy tensor of the fluid and the electromagnetic field together. By neglecting the dissipation due to the viscosity, and the heat conduction in the fluid frame $T^{\mu \nu}$ reads as \cite{1989rfmw.book.....A},

\begin{align}
T^{\nu \mu}=\left(w+|b|^{2}\right) u^{\nu} u^{\mu}+\left(p+\frac{1}{2} |b|^{2}\right) g^{\nu \mu}-b^{\nu} b^{\mu},
\end{align}
where $p$ is the gas pressure, $|b|^2=2p_{\rm m}$ where $p_m$ is the magnetic pressure in the fluid \cite{1989rfmw.book.....A}, and $\leftidx{^*}{F}{^{\nu \mu}}$ is the Hodge dual Faraday tensor

\begin{equation}
    \leftidx{^*}{F}{^{\nu \mu}}=b^{\nu} u^{\mu}-b^{\mu} u^{\nu},
\end{equation}
where $b^{\mu}$ is the four-vector magnetic field. We proceed here following \cite{Komissarov_2006}, by the assumption of axially symmetric and stationary space-time

\begin{align}
u^r=u^{\theta}=b^r=b^{\theta}=0.
\end{align}
It is clear that considering these assumptions, the only nontrivial result follows from solving the stress-energy conservation. By choosing to have $\Omega=\Omega(\ell)$ as the integrability condition, and projection of the conservation of stress-energy tensor into the plane normal to four-velocity, we obtain \cite{1974AcA....24...45A,Komissarov_2006}

\begin{align}\label{A7}
\frac{1}{w}\nabla_{i}p=-\nabla_{i}\ln u_{t}+\frac{\Omega\nabla{}_{i}\ell}{1-\Omega\ell}-\frac{\nabla_{i}\Tilde{p}_{\rm m}}{\Tilde{w}},
\end{align}
where $\Tilde{p}_{\rm m}=\mathcal{L}p_{\rm m}$, $\Tilde{w}=\mathcal{L}w$, and in the static set up $\mathcal{L}=-g_{t t} g_{\phi \phi}$. Then this equation turns easily to

\begin{align}\label{9}
    \int_0^{p}{\frac{{\rm d}p}{w}}+\int_0^{\Tilde{p}_{\rm m}}{\frac{{\rm d}\Tilde{p}_{\rm m}}{\Tilde{w}}}&=-\ln{|u_t|}+\ln{|(u_t)_{\rm in}|}+\int_{\ell_{\rm in}}^{\ell}{\frac{\Omega{\rm d}\ell}{1-\Omega\ell}}.
\end{align}
The constant of integration was chosen so that the pressures vanished on the surface of the disk and its inner edge i.e., $u_t=(u_t)_{\rm in}$ and $\ell=\ell_{\rm in}$.

Adopting \cite{Komissarov_2006}, to be able to express the integrals of equation \eqref{9}
in terms of elementary functions, we need to assume these extra relations

\begin{align}\label{state}
p=K w^{\kappa}, \quad \tilde{p}_{\mathrm{m}}=K_{\mathrm{m}} \tilde{w}^{\eta}
\end{align}
where $K$, $\kappa$, $K_{\mathrm{m}}$ and $\eta$ are constants. By this particular choice for equations of state, the Von Zeipel theorem is fulfilled. Therefore the equation \eqref{9} is fully integrated, and we obtain

\begin{align}\label{eq:FinalEq}
W-W_{\mathrm{in}}+\frac{\kappa}{\kappa-1}\frac{p}{w}+\frac{\eta}{\eta-1} \frac{p_m}{w}=\int_{\ell_{\rm in}}^{\ell}{\frac{\Omega{\rm d}\ell}{1-\Omega\ell}},
\end{align}
where $W=\ln|u_t|$.
The model parameters to have a unique solution are $\kappa>0$, $\eta$, $W_{\rm in}$, $w_{\rm c}$ enthalpy at the center, and the magnetization parameter $\beta_{\rm c}>0$, which is the ratio of the gas pressure to magnetic pressure, at the center. The positivity of $\kappa$ and $\beta_{\rm c}$ is due to avoiding divergence in the equation \eqref{eq:FinalEq}. Also, one needs to specify $\ell(r,\theta)$ to fix the geometry of the equipotential surfaces, where we consider different cases in what follows.


\subsection{ Constant angular momentum}
\label{3}
The first angular momentum case we have considered in this work has the constant distribution profile $\ell=\ell_0$. Therefore, the right-hand side of the equation \eqref{eq:FinalEq} vanishes

\begin{align}\label{eq:FinalEqlconst}
W-W_{\mathrm{in}}+\frac{\kappa}{\kappa-1}\frac{p}{w}+\frac{\eta}{\eta-1} \frac{p_{\rm m}}{w}=0.
\end{align}
The disk centre is found at the larger radius where $\ell_0$ intersects with the local Keplerian angular momentum. In contrast, $\ell_0$ is larger than the radius of the marginally stable Keplerian orbit $\ell_{\rm ms}$. One can find the cusp point at the radius of the intersection of the specific angular momentum and the Keplerian one \cite{1974AcA....24...45A,Komissarov_2006}.
Therefore, the disk surface is fully determined by choice of $W_{\rm in}$ independently of the magnetic field \cite{1989PASJ...41..133O}. In this case, the total potential reads as

\begin{align}\label{K33}
    W(r,\theta)=\frac{1}{2}\ln|\frac{\mathcal{L}}{\mathcal{A}}|,
\end{align}
where $\mathcal{A}=g_{\phi\phi}+\ell^2_0g_{tt}$ in the static case. Thus $W$ satisfies this relation \cite{1978A&A....63..221A},

\begin{align}
    \left\{
  \begin{array}{@{}ll@{}}
  W_{\rm in}\leq W_{\rm cusp} & \text{if}\ |\ell_{\rm ms}|<|\ell_0|< |\ell_{\rm mb}|, \\
  W_{\rm in}<0 & \text{if}\ |\ell_0|\geq |\ell_{\rm mb}|.
    \end{array}\right.
\end{align}
Also, the gas pressure at the center $p_c$, reads as

\begin{align}
    p_{\rm c}= w_{\rm c}(W_{\rm in}-W_{\rm c})\left(\frac{\kappa}{\kappa-1}+\frac{\eta}{\beta_{\rm m_c}(\eta-1)}\right)^{-1},
\end{align}
where the subscript $\rm c$ refers to the mentioned quantity at the center. Also, $\betac=p_{\rm c}/p_{\rm mc}$ is the magnetization parameter at the center.
The variables of model are then $W$, $w$, $p$, $p_{\rm m}$, $u^t$, $u^{\phi}$, $b^t$ and $u^{\phi}$. So by using equation of state, one can find $K$ and $K_{\rm m}$, then the solution is easily obtained utilizing \eqref{eq:FinalEq} and \eqref{K33} \cite{Komissarov_2006}.


\subsection{\label{5w} Power-low angular momentum distribution}
In the second case, we relax the condition of constant angular momentum. However, by considering pure rotation and a barotropic equation of state, angular velocity can state as a function of specific angular momentum $\Omega=\Omega(\ell)$. In this case, following \cite{2015MNRAS.447.3593W} we consider   
\begin{align}
\Omega(\ell)=c\ell^{n},    
\end{align}
where $c$ and $n$ are constant parameters. Then the equation \eqref{eq:FinalEq} can be written as \cite{2015MNRAS.447.3593W},

\begin{align}\label{eq:FinalEqw}
&W-W_{\mathrm{in}}+\frac{\kappa}{\kappa-1}\frac{p}{w}+\frac{\eta}{\eta-1} \frac{p_{\rm m}}{w}\\
&=\frac{1}{n+1}\ln\left(\frac{c\ell_{\rm in}^{n+1}-1}{c\ell^{n+1}-1}\right).\nonumber\
\end{align}
Of course, one needs to calculate $c$ and $n$ to obtain angular momentum distribution. It can be done using the center of the torus and the cusp of the disk. In the next step, by calculating $\Omega(r_{\rm c})$, $\Omega(r_{\rm cusp})$, and using the original definition of $\Omega$ in equation \eqref{omel}, one obtains a system of equations that can be solved analytically to find parameters $c$ and $n$ and the distribution \cite{2015MNRAS.447.3593W}. 

At this stage, we have all we need to build magnetized thick disks. We can find the gas and magnetic distribution, the enthalpy distribution, and the rest-mass density distribution \eqref{state}. The fluid enthalpy is giving by $w=\rho h$, where $h = 1 + \epsilon + \frac{p}{\rho}$ is the specific enthalpy which in this case it can approximate by $h \sim1 $. This approximation is supported by the fact that the fluid is not relativistic from a thermodynamical point of view, and the internal energy and the pressure are very small compared to the rest-mass density. However, relaxing this assumption does not lead to noticeable differences in the models we considered. It is worth noticing that for a circular rotating perfect fluid, the shapes and location of the equipressure surface $p(r, \theta)$ = const. are characterized by the assumed angular momentum distribution independently of the equation of state and the assumed entropy distribution \cite{1980AcA....30....1J}.


\subsection{\label{5q} Trigonometric  angular momentum distribution}

The third case is also considered for non-constant angular momentum where presented in \citep{2009A&A...498..471Q}. This is a reasonable assumption from a physical point of view to combining constant profile with the Keplerian one. Following them, we have assumed an angular momentum distribution for the hydrodynamical case given by

\begin{align}
 \ell(r,\theta)=
   \left\{
  \begin{array}{@{}ll@{}}
  \ell_0\left(\frac{\ell_{\rm K(r)}}{\ell_0}\right)^{\alpha}\sin^{2\delta}, & r\geq r_{\rm ms}, \\
     \ell_0(\zeta)^{-\alpha}\sin^{2\delta}, & r<r_{\rm ms},
    \end{array}\right.
\end{align}
where $\ell_0=\zeta \ell_{\rm K}(r_{\rm ms})$, and $\ell_{\rm K}$ is the Keplerian angular momentum in the equatorial plane, and

\begin{align}
   0\leq \alpha\leq 1, \quad -1\leq \delta \leq 1, \quad -1\leq \zeta\leq \frac{\ell_{\rm K}(x_{\rm mb})}{\ell_{\rm K}(x_{\rm ms})}.
\end{align}
In this case, the equipressure surface which starts from the cusp is the marginally bound for $\alpha=0$, $\delta=0$, and $\zeta=\ell_K(x_{\rm mb})/\ell_{\rm K}(x_{\rm ms})$. Also, for the MHD case

\begin{align}
 \ell(r,\theta)=
   \left\{
  \begin{array}{@{}ll@{}}
      \ell_0\left(\frac{\ell_K(r)}{\ell_0}\right)^{\alpha}\sin^{2\delta}, & r\geq r_{\rm ms}, \\
\\
       \ell_{\rm ms}(r)\sin^{2\delta}, & r<r_{\rm ms},
    \end{array}\right.
\end{align}
where $\ell_{\rm ms}(r)$ is calculated on the equatorial plane via considering $\Omega_{\rm ms}$ simply by using equation \eqref{omel}.

This procedure is described as follows. First, one should substitute $i=r$ in equation \eqref{A7}, and after replace $i=\theta$ in this equation. Second, divide the two equations to obtain \citep{2009A&A...498..471Q},

\begin{align}
    \frac{\partial_rp}{\partial_{\theta}p}=\frac{\partial_rg^{tt}+\ell^2\partial_rg^{\phi\phi}}{\partial_{\theta}g^{tt}+\ell^2\partial_{\theta}g^{\phi\phi}}:= -F(r,\theta)
\end{align}
Thus, for a given angular momentum distribution $\ell$, the function $F$ is known. In addition, considering $\theta=\theta(r)$ as the explicit equation for the equipressure surface leads to %

\begin{align}
    \frac{d\theta}{dr}=F(r,\theta).
\end{align}
One can solve this equation for different initial conditions. Therefore one can obtain all the possible locations for the equipressure surfaces.



In this work, we choose to construct this magnetized thick disk in various cases in the distorted Schwarzschild space-time, where in the next section, this space-time is briefly explained.

\section{Distorted Schwarzschild black hole}\label{sec3}
The Schwarzschild space-time is the vacuum, regular, unique static solution of Einstein's field equation, which is asymptotically flat. Another related solution that is not asymptotically flat is obtained by assuming an external distribution of matter outside the horizon \cite{1982JMP....23..680G,Chandrasekhar:579245}.

In fact, the exterior of the slowly rotating compact object, in the presence of an external static and axially symmetric matter distribution, can be described by the distorted Schwarzschild black hole.
To describe this space-time, we start with the Weyl metric, which is a convenient choice for describing any axially symmetric and stationary space-time. In the static case, this metric reads as

\begin{align}\label{GHmetric}
\dd s^2 & =-e^{2\psi}\dd t^2 +e^{2(\gamma-\psi)}(\dd\rho^2+\dd z^2) +e^{-2\psi}\rho^2 \dd\phi^2\,.
\end{align}
There are two metric functions in the Weyl metric, namely $\psi=\psi(\rho,z)$ and $\gamma=\gamma(\rho,z)$. The function $\psi$ plays the role of gravitational potential and in flat space

\begin{align}
 ds^2=d\rho^2+dz^2+\rho^2 d\phi^2.  
\end{align}
The function $\psi$ obeys the Laplace equation, which is the key factor to construct such a solution \cite{1982JMP....23..680G,Chandrasekhar:579245},

\begin{align}
\psi_{s_,\rho\rho}+\frac{1}{\rho}\psi_{s_,\rho}+\psi_{s_,zz}=0\label{laplap}.
\end{align}
This equation is also the integrability condition for the metric function $\gamma$, and it is obtained by the explicit form of the function $\psi$ as follows 

\begin{align}\label{gammaeq}
\gamma_{,\rho}=&\rho(\psi^2_{,\rho}-\psi^2_{,z}),\nonumber\\
\gamma_{,z}=&2\rho_{}\psi_{,\rho}\psi_{,z}.\
\end{align}
The relation between Schwarzschild coordinates and Weyl coordinates is given by 

\begin{align}
\rho=\sqrt{r(r-2M)}\sin{\theta} \quad z=(r-M)\cos{\theta},
\end{align}
where $M$ is a parameter that is considered as the mass of the compact object. However, it is more appropriate to rewrite this metric in prolate spheroidal coordinate $(t, x, y, \phi)$, via this transformation relation

\begin{align}
    {x} & =\frac{1}{2{M}}(\sqrt{\rho^2+({z}+{M})^2}+\sqrt{\rho^2+({z}-{M})^2}),\\
    {y} & =\frac{1}{2{M}}(\sqrt{\rho^2+({z}+{M})^2}-\sqrt{\rho^2+({z}-{M})^2}) \nonumber.
\end{align}

The relation to Schwarzschild coordinates is given by
\begin{align}\label{transf1}
x & =\frac{r}{M}-1,\\
 y & = \cos\theta \nonumber.
\end{align}
In these coordinates, the metric functions for Schwarzschild solution $\psi_s$ and $\gamma_s$ are written as 

\begin{align}
     \psi_s & =\frac{1}{2}\ln{\frac{{x}-1}{{x}+1}}\,,\\
     \gamma_s & =\frac{1}{2}\ln{\frac{{x}^2-1}{{x}^2-{y}^2}}\nonumber\,.
\end{align}
By replacing these two functions, ultimately, the metric \eqref{GHmetric} takes this form 


\begin{align}\label{Dmetric}
	\dd s^2 = &- \left( \frac{x-1}{x+1} \right) e^{2\hat{\psi}} \dd t^2+ M^2 (x+1)^2 e^{-2\hat{\psi}} \\
	& \left[e^{2\hat{\gamma}} \left( \frac{\dd x^2}{x^2-1} + \frac{\dd y^2}{1-y^2} \right) + (1-y^2) \dd{\phi}^2 \right]\nonumber \,,
\end{align}
where $t \in (-\infty, +\infty)$, $x \in (1, +\infty)$, $y \in [-1,1]$, and $\phi \in [0, 2\pi)$. In this metric, the location of the horizon is at $x=1$ and the singularity is at  and $x=-1$.

The general solution for the metric functions $\hat{\psi}$ and $\hat{\gamma}$ can be expressed in terms of Legendre polynomials \cite{Chandrasekhar:579245,1997PhLA..230....7B}, where the expansion coefficients $a_n$, are called multipole moments due to an external field. For $a_n=0$, we have no moment and, of course, the Schwarzschild space-time. The next one $a_1$ is called the dipole moment, and $a_2$ is the quadrupole moment, and so on. Some conditions should be satisfied \cite{1982JMP....23..680G}, in particular, to fulfil this requirement of having no singularities on the symmetry axis, we have to demand



\begin{equation}\label{equi}
\sum_{n=1} a_{2n-1} = 0 \,.
\end{equation}
In the following, we will restrict our attention to a distortion up to the relatively small quadrupole moment, assuming contributions of higher orders can be neglected. Moreover, for simplicity, we adapted the notation $\rm q$ for the quadrupole moment instead of $a_2$. Therefore these functions up to the quadrupole moment are expressed as follows

\begin{align}
\hat{\psi} & = -\frac{\rm q}{2}\left[-3x^2y^2+x^2+y^2-1\right],\\
\hat{\gamma} & = -2x{\rm q}(1-y^2)\nonumber\\
  &+\frac{{\rm q}^2}{4}(x^2-1)(1-y^2)(-9x^2y^2+x^2+y^2-1).\
  \end{align}
One can also derive the various physical quantities from the metric in this space-time, including Keplerian angular momentum that we need to construct this thick disk model.

\begin{align}
\ell_{\rm K}=\sqrt{\frac{1+qx-qx^3}{1+qx+qx^2}} \frac{(x+1)^{\frac{3}{2}}}{x-1}Me^{-2\hat{\psi}}.
\end{align}
This is worth mentioning that similar to the context of Newtonian set up, one can think of positive quadrupoles as the distortion of the black hole due to an oblate configuration of external distribution of matter, and negative quadrupoles as a prolate mass distribution, along the symmetry axis. Therefore, we expect to have more radially extended disk for positive values of $\rm q$, and more vertically for negative quadrupoles choices in comparison. This structure is expected due to both the gravitational forces from the central object and the one for the external source where is closer to the disk, in each case.


In the following section, we present the results and discussion of the disk configuration in this space-time.
\section{Result and discussion}
\label{resultsd}
This section presents various results showing the effect of the parameter $\rm q$ related to the distortion of the black hole in the thick disk model. We start by considering constant angular momentum; then, we continue by discussing the non-constant cases.

\subsection{\label{constant}discussion on the constant angular momentum}

\begin{figure}
    \centering
    \includegraphics[width=\hsize]{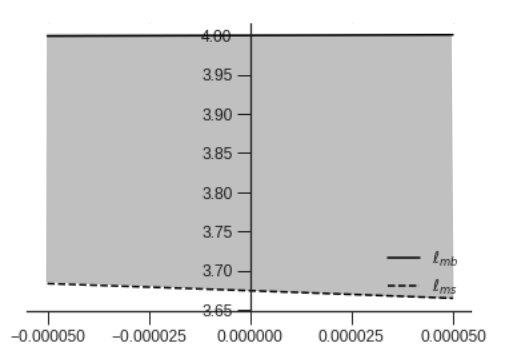}
    \caption{profiles of $\ell_{\rm ms}$ and $\ell_{\rm mb}$ as a function of $\rm q$.}
    \label{fig:1}
\end{figure}

In this subsection, we assume specific angular momentum is constant and fixed to be $\ell(x,y)=\ell_0$. As mentioned in section \ref{3}, different disks can be constructed upon different choices of the value of $\ell_0$. In this model, the different regions are bounded by the value of the angular momentum at the marginally bound orbit $\ell_{\rm mb}$, and at the marginally stable circular orbit $\ell_{\rm ms}$. Our interest is in the region where we can construct disks with or without a cusp, i.e., $\ell_{\rm ms}<\ell_0<\ell_{\rm mb}$. Thus, our first interrogation is about the variation of $\ell_{\rm ms}$ and $\ell_{\rm mb}$ as a function of $\rm q$ where depicted in Figure \ref{fig:1} for $\rm q\in [-0.0001,0.0001]$. We see that the specific angular momentum at the marginally stable orbit is decreasing. Consequently, the area where closed equipotential surfaces are possible becomes larger for larger values of $\rm q$. This analysis shows how to choose $\ell_0$ to have a solution for which closed equipotential surfaces can exist. 

The next step is to find the location of the cusp, $\xcusp$, and the centre, $\xc$, which are chosen by fixing $\ell_0$. These are the points that we have the Keplerian angular momentum equals to the chosen angular momentum $\ell _0$ at the equatorial plane, 
\begin{equation}
    \ell_0=\ell_K(x) \quad  \text{at} \quad y=0.
\end{equation}
In this work, we chose the angular momentum to be $\ell_0=3.77$, which is compatible with Figure \ref{fig:1}. Therefore, it allows us to have the solution with a cusp for all values of $\rm q$. 

\begin{figure*}[!htbp]
   \centering
\begin{tabular}{ccc}
\includegraphics[scale=0.4]{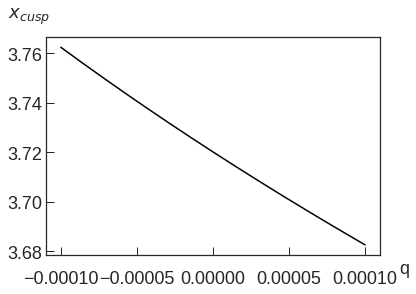}&
\includegraphics[scale=0.4]{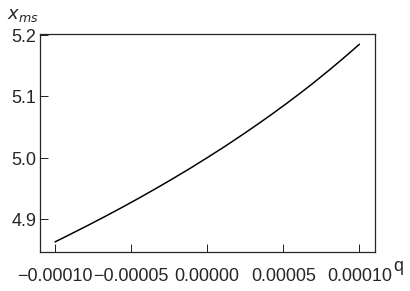}&
\includegraphics[scale=0.4]{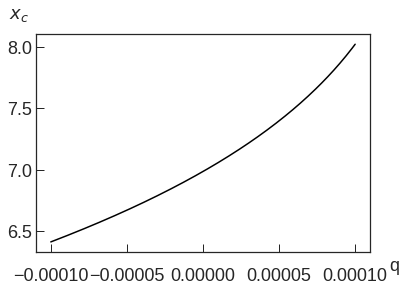}
\end{tabular}
    \caption{Variation of $\xcusp$, $x_{\rm ms}$, and $\xc$ as a function of $\rm q$ for a fixed value of specific angular momentum, respectively.}
    \label{fig:2} 
\end{figure*}
\begin{figure*}[!htbp]
  \centering
 \includegraphics[width=17cm]{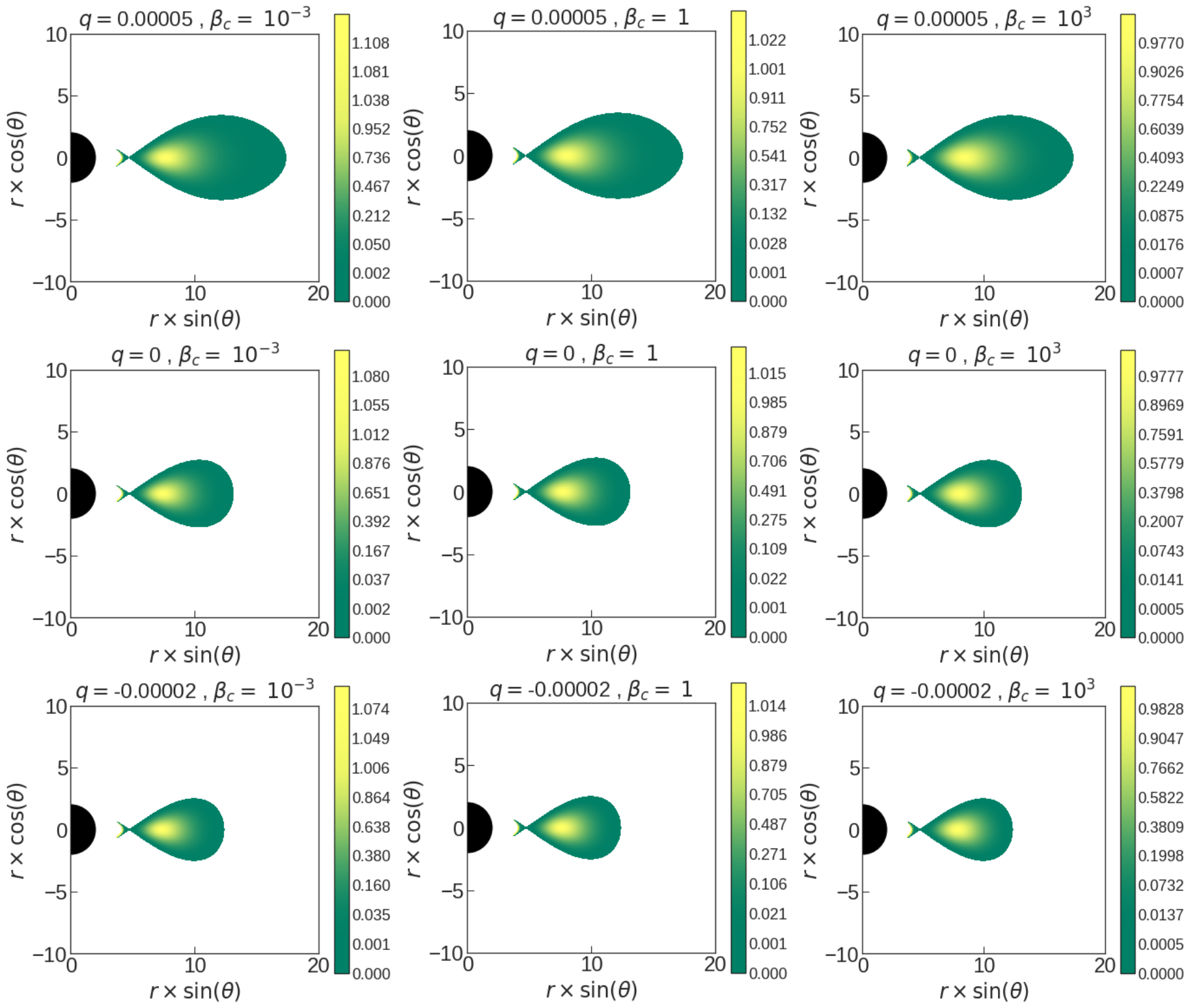}
 \caption{The rest-mass density distribution for constant angular momentum. The models $\rm A$, $\rm C$ and $\rm E$ of the Table \ref{tab:1} are presented in rows. The index $1$, $2$ and $3$ of each model correspond to each column from left to right.} 
    \label{fig:5} 
 \end{figure*}

In Figure \ref{fig:2}, variation of $x_{\rm cusp}$, $x_{\rm ms}$ and $x_{\rm c}$ over $\rm q$ is plotted. As we see, $\xcusp$ is a monotone decreasing function of $\rm q$. Simultaneously, the marginally stable location $x_{\rm ms}$ and $x_{\rm c}$ are monotone increasing function of $\rm q$. Therefore, $\xcusp$ and $x_{\rm ms}$ are pushing further from each other, confirming the extension of the area where a solution with closed equipotential surfaces can be built in Figure \ref{fig:1}. This Figure also indicates that as the value of $\rm q$ increases, the distance between the cusp and the centre of the disk increases, and we would expect to have a more radially expanded disk as we see in Figure \ref{fig:5}.

\begin{figure}[!htbp]
   \centering
\begin{tabular}{c}
\includegraphics[width=7.5cm]{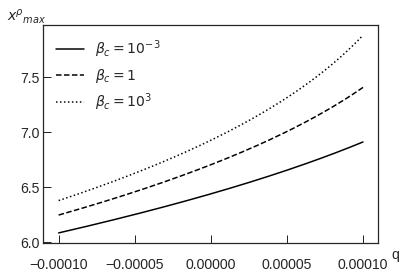}\\
\includegraphics[width=7.5cm]{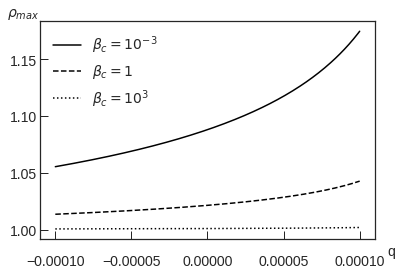}
\end{tabular}
    \caption{Amplitude and location of the rest-mass density maximum over $\rm q$, for $\ell(x,y)=\ell_0=3.83$.}
    \label{fig:3} 
\end{figure}

\begin{figure}[!htbp]
   \centering
\begin{tabular}{c}
\includegraphics[width=7.5cm]{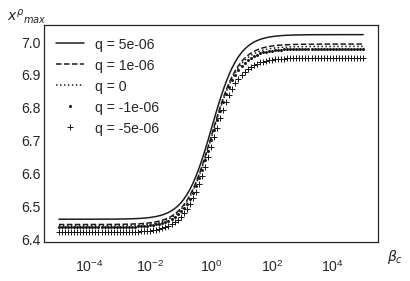}\\
\includegraphics[width=7.5cm]{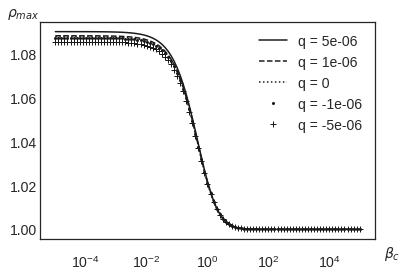}
\end{tabular}
    \caption{Amplitude and location of the rest-mass density maximum over $\rm \betac$, and the five values of $\rm q$ described in Table \ref{tab:1}. This is plotted for $\ell(x,y)=\ell_0=3.83$.}
    \label{fig:4} 
\end{figure}

Figure \ref{fig:3} shows the profile of the maximum rest-mass density and its location in the equatorial plane over $\rm q$, for different values of $\beta_{\rm c} $\footnote{The morphology of the solutions for nondistorted case indicates that there is no significant change in the solutions for magnetization values above $\beta_{\rm c}=10^{3}$ and below $\beta_{\rm c}=10^{-3}$.}. We see that for a fixed value of $\rm q$, the rest-mass density maximum is a decreasing function, and its location is an increasing function of $\beta_{\rm c}$. It is worth mentioning that the lower the value of $\betac$, the stronger the magnetization effect by its definition. Considering the assumptions of MHD, the magnetic field causes the rest-mass density increase in the inner part of the disk as a response to the magnetic pressure, and as a result, its place shifted farther away accordingly. This is compatible with the proceeding literature. Furthermore, for a fixed $\beta_{\rm c}$, the maximum and its location are an increasing function of $\rm q$. In fact, the disk has been constructed between the compact object and the mass distribution parameterized by $\rm q$ in its vicinity. Therefore, as the parameter, $\rm q$ increases, the effect of this external source is more manifest, and maximum density becomes higher at the radios farther away from the central object and closer to the external source. Then for some proper combination of parameters, $\rm q$ and $\beta_{\rm c}, $ can compensate some of their effects. In Figure \ref{fig:4}, the location of the rest-mass density maximum and its amplitude are plotted as a function of $\beta_{\rm c}$. The results are compatible with the result of Figure \ref{fig:3}.


One can construct various disk models with different shapes due to different cusp locations and the centre. We have considered models denoted by the letters $\rm A$, $\rm B$, $\rm C$, $\rm D$, and $\rm E$, to show the effect of the quadrupole $\rm q$. Also, by subscripts $1,2,3$ to show the effect of the magnetization parameter $\rm \betac$ on the disk's shape. The parameters of each case are given in Table \ref{tab:1}. All these solutions are depicted in Figure \ref{fig:4}.

In Figure \ref{fig:5} we have examined models $\rm A$, $\rm C$ and $\rm E$ of the Table \ref{tab:1}. In the first column the magnetisation parameter is set to be $\rm \beta_c=10^{-3}$, in the second column is $\betac=1$, and in the third column $\betac=10^{3}$ to compare with Figure \ref{fig:3}. Also, rows correspond to different quadrupole values. Figure \ref{fig:5}, shows that $\rm q$ and $\betac$ both have significant effects on the structure of the thick disk model. In fact,  magnetic pressure does not touch the overall shape of the disk and its radial expansion strongly. However, further analysis shows a slight change occurs in the range of isodensity contours. This is compatible with increasing density toward the inner part of the disk. 

On the other hand, the role of $\rm q$ is more substantial on the shape of the disk and its place. In fact, as the value of $\rm q$ increases, the disk has more radially extended and shifted away from the central object. This is also well matched with the result of Figures \ref{fig:3} and \ref{fig:4}.

\begin{table}[!htbp]
    \centering
\begin{tabular}{p{0.6cm}p{1.7cm}p{.8cm}p{.8cm}p{.8cm}p{.8cm}p{0.8cm}}
\hline
 Case   & $\rm q$ & $\xin$ & $\xc$ & $\xmaxP$ & $\xmaxPM$ & $\betac$   \\
\hline
 \hline
  $\rm B_1$  & $0.000001$ & $3.719$ & $6.991$ & $6.439$ & $6.569$ & $10^{-3}$ \\
 \hline
 $\rm B_2$  & $0.000001$ & $3.719$ & $6.991$ & $6.710$ & $6.849$ & $1$ \\
 \hline
 $\rm B_3$  & $0.000001$ & $3.719$ & $6.991$ & $6.990$ & $7.139$ & $10^{3}$ \\
 \hline
 \hline
  $\rm B_1$  & $0.0000001$ & $3.720$ & $6.985$ & $6.440$ & $6.570$ & $10^{-3}$ \\
 \hline
 $\rm B_2$  & $0.0000001$ & $3.720$ & $6.985$ & $6.700$ & $6.840$ & $1$ \\
 \hline
 $\rm B_3$  & $0.0000001$ & $3.720$ & $6.985$ & $6.980$ & $7.130$ & $10^{3}$ \\
 \hline
 \hline
  $\rm C_1$  & $0$ & $3.720$ & $6.984$ & $6.440$ & $6.570$ & $10^{-3}$ \\
 \hline
 $\rm C_2$ & $0$ & $3.720$ & $6.984$ & $6.700$ & $6.840$ & $1$ \\
 \hline
 $\rm C_3$ & $0$ & $3.720$ & $6.984$ & $6.980$ & $7.130$ & $10^{3}$ \\
 \hline
 \hline
  $\rm D_1$  & $-0.0000001$ & $3.720$ & $6.983$ & $6.440$ & $6.570$ & $10^{-3}$ \\
 \hline
 $\rm D_2$  & $-0.0000001$ & $3.720$ & $6.983$ & $6.700$ & $6.840$ & $1$ \\
  \hline
 $\rm D_3$  & $-0.0000001$ & $3.720$ & $6.983$ & $6.980$ & $7.130$ & $10^{3}$ \\
 \hline
 \hline
$\rm D_1$  & $-0.000001$ & $3.721$ & $6.977$ & $6.430$ & $6.560$ & $10^{-3}$ \\
 \hline
 $\rm D_2$  & $-0.000001$ & $3.721$ & $6.977$ & $6.700$ & $6.830$ & $1$ \\
  \hline
 $\rm D_3$  & $-0.000001$ & $3.721$ & $6.977$ & $6.970$ & $7.120$ & $10^{3}$ \\
 \hline
\end{tabular}
    \caption{Parameters of the considered solutions for a constant specific angular moment $\ell_0=3.77$.}
    \label{tab:1}
\end{table}




\begin{figure}[!htbp]
   \centering
\includegraphics[width=8cm]{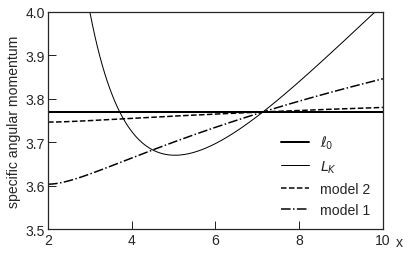}
\caption{The models $1$, $2$ are radial profiles of the specific angular momentum in the equatorial plane with different slopes. The constant specific angular momentum is shown with a thick straight line, which is chosen to be $\ell(x,y)=\ell_0=3.83$. This profile corresponds to $\rm q=0.00002$.}
\label{fig:6} 
\end{figure}


\begin{figure}[!htbp]
   \centering
\begin{tabular}{c}
\includegraphics[width=7cm]{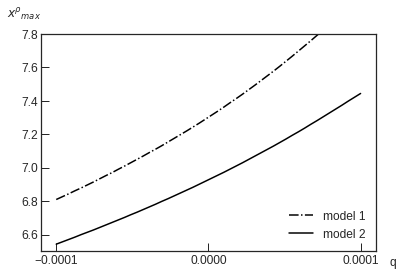}\\
 \includegraphics[width=7cm]{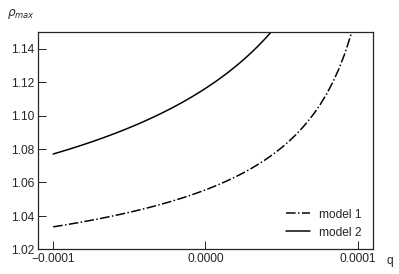}
\end{tabular}
    \caption{Amplitude and location of the rest-mass density maximum over $\rm q$. The different curves represent different distributions of specific angular momentum depicted in the Figure \ref{fig:6}.}
    \label{fig:7}  
\end{figure}

\begin{figure}[!htbp]
   \centering
   \begin{tabular}{c}
\includegraphics[width=8cm]{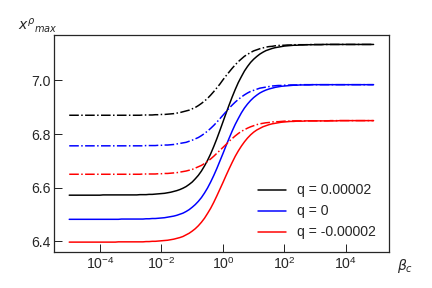}\\
\includegraphics[width=8cm]{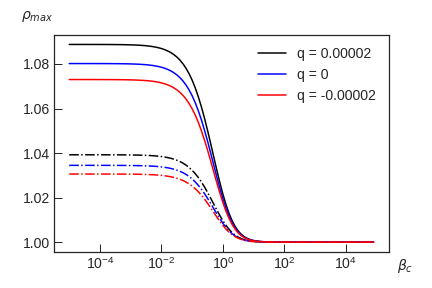}
\end{tabular}
    \caption{Amplitude and location of the rest-mass density maximum as a function of $\rm \betac$. The solid lines correspond to the model $2$ and the dot-dashed lines present the model $1$.}
    \label{fig:8}  
\end{figure}


\subsection{discussion on power-law distribution}

In this section, we discuss different profiles of $\ell(x,y)$ and values of parameter $\rm q$ following the method described in \cite{2015MNRAS.447.3593W} which is described in Subsection \ref{5w}. This method requires choosing a fixed specific angular momentum $\rm \ell_0$ to fix the centre $\xc$ and cusp $\xcusp$. This procedure proceeds as follows. We keep $\xc$ fixed and set a new cusp point called $\xcusp^{\prime}$, choosing between $\xcusp$ and $x_{\rm ms}$. The slope of each profile then depends on the chosen value for $\xcusp^{\prime}$. Therefore, we chose two different values to get two $\ell(x,y)$ distributions denoted by a numeral index $1$ and $2$. In Figure \ref{fig:6}, the profile of the specific angular momentum distribution is plotted for $\rm q=0.00002$. 

In addition, as in the previous case, $\xcusp$ and $\xc$ vary with quadrupole's value, then it is expected the slope changes also with $\rm q$.



In Figures \ref{fig:7} and \ref{fig:8}, we turned on the magnetic field, and we studied the effect of magnetic pressure combines with different distributions of angular momentum presented in Figure \ref{fig:6}. In Figure \ref{fig:7} a global view of the variation of the location and amplitude of the maximum rest-mass density over $\rm q$ is plotted. In addition, Figure \ref{fig:8} shows the maximum rest-mass density profile and its value in the equatorial plane over magnetisation parameter for different values of quadrupole and two models. For $\beta_{\rm c}$ and $\rm q$, the same behaviour as the constant case applies, so we expect the more radially extended disk structure for higher values of $\rm q$. Increasing in $\beta_{\rm c}$ decreases the rest-mass density maximum and places its location farther away. Also, an increase in $\rm q$ has the same impact as $\beta_{\rm c}$ on the rest-mass maximum and its location. Also, these Figures suggest the effect of $\rm q$ is more intense for model $1$, which has a steeper slope, while the impact of $\beta_{\rm c}$ is stronger on model $2$, which is closer to the Keplerian one. Also, a further inspection reveals that when $\rm q$ is higher, the influence of changing in $\beta_{\rm c}$ is also greater for all models.



%

\begin{table}
    \centering
\begin{tabular}{p{0.7cm}p{1.7cm}p{0.7cm}p{0.8cm}p{.8cm}p{.8cm}}
\hline
 Case   & $\rm q$ & $\xin$ & $\xc$ & $\xmaxP$ & $\xmaxPM$   \\
    \hline
    \hline
$\rm F_1$   & $0.000050$ &  $4.5$ & $7.206$ & $7.360$ & $7.469$  \\
 \hline
  $\rm F_2$   & $0.000050$ & $3.8$ & $7.206$ & $6.957$ & $7.147$  \\
  \hline
  \hline
  $\rm G_1$   & $0.000020$ & $4.5$ & $7.241$ & $7.162$ & $7.249$  \\
 \hline
  $\rm G_2$   & $0.000020$ & $3.8$ & $7.241$ & $6.810$ & $6.971$  \\
 \hline
 \hline
  $\rm H_1$   & $0$ & $4.5$ & $7.353$ & $7.044$ & $7.120$  \\
 \hline
  $\rm H_2$   & $0$ & $3.8$ & $7.353$ & $6.719$ & $6.866$  \\
 \hline
 \hline
  $\rm I_1$   & $-0.000015$ & $4.5$ & $7.519$ & $6.961$ & $7.030$  \\
 \hline
  $\rm I_2$   & $-0.000015$ & $3.8$ & $7.519$ & $6.655$ & $6.792$   \\
 \hline
  \hline
 $\rm J_1$   & $-0.000020$ & $4.5$ & $7.817$ & $6.934$ & $7.001$ \\
 \hline
  $\rm J_2$   & $-0.000020$ & $3.8$ & $7.817$ & $6.634$ & $6.768$  \\
 \hline
\end{tabular}
    \caption{Parameters of the considered solutions for power-law angular momentum distribution. The alphabetical enumeration of the models $F,G,H,I$ and $J$ refers to the different values of $\rm q$ given in Table \ref{tab:2}. The indexes in each model refer to the different specific angular momentum distributions in Figure \ref{fig:6}. All the models have $\beta_c=10^{-3}$. }
    \label{tab:2}
\end{table}

\begin{table}
    \centering
\begin{tabular}{p{0.7cm}p{1.7cm}p{0.7cm}p{0.8cm}p{.8cm}p{.8cm}}
\hline
 Case   & $\rm q$ & $\xin$ & $\xc$ & $\xmaxP$ & $\xmaxPM$   \\
    \hline
    \hline
$\rm F_1$   & $0.000050$ &  $4.5$ & $7.817$ & $7.816$ & $7.939$  \\
 \hline
  $\rm F_2$   & $0.000050$ & $3.8$ & $7.817$ & $7.816$ & $8.076$  \\
  \hline
  \hline
  $\rm G_1$   & $0.000020$ & $4.5$ & $7.519$ & $7.518$ & $7.611$  \\
 \hline
  $\rm G_2$   & $0.000020$ & $3.8$ & $7.519$ & $7.518$ & $7.720$  \\
 \hline
 \hline
  $\rm H_1$   & $0$ & $4.5$ & $7.353$ & $7.353$ & $7.431$  \\
 \hline
  $\rm H_2$   & $0$ & $3.8$ & $7.353$ & $7.352$ & $7.528$  \\
 \hline
 \hline
  $\rm I_1$   & $-0.000015$ & $4.5$ & $7.241$ & $7.241$ & $7.311$  \\
 \hline
  $\rm I_2$   & $-0.000015$ & $3.8$ & $7.241$ & $7.240$ & $7.400$   \\
 \hline
  \hline
 $\rm J_1$   & $-0.000020$ & $4.5$ & $7.206$ & $7.205$ & $7.274$ \\
 \hline
  $\rm J_2$   & $-0.000020$ & $3.8$ & $7.206$ & $7.205$ & $7.360$  \\
 \hline
\end{tabular}
    \caption{The same cases as table \ref{tab:2} for $\beta_c=10^{3}$. }
    \label{tab:3}
\end{table}

       \begin{figure}[!htbp]
   \centering
\includegraphics[width=\hsize]{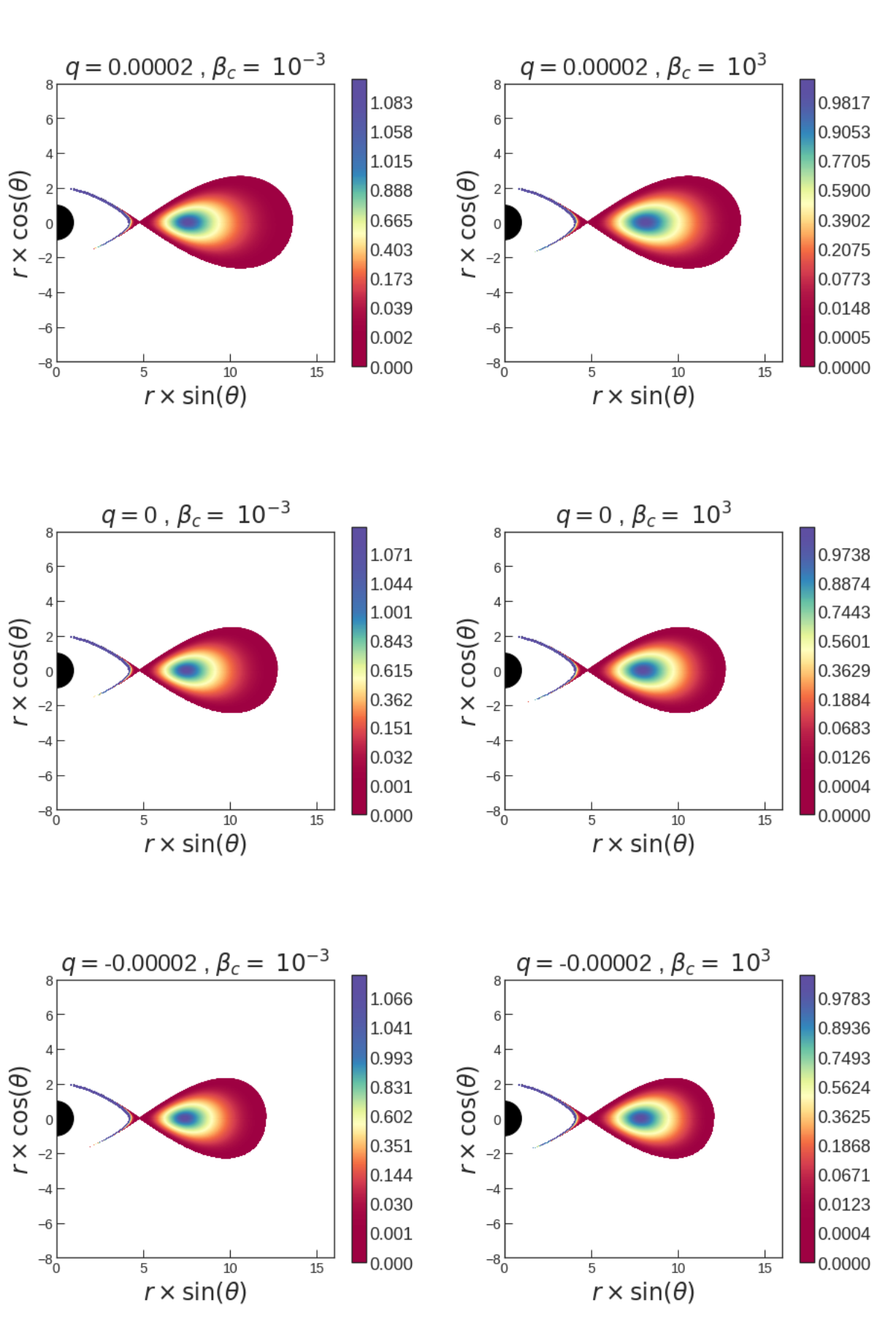}
    \caption{Map of the rest-mass density distribution for the angular momentum model $2$ as described in the Figure \ref{fig:6}. Each row corresponds to different values of $\rm q$, and each column shows different value of $\betac$.}
    \label{fig:9} 
  \end{figure}

    \begin{figure}[!htbp]
   \centering
\includegraphics[width=\hsize]{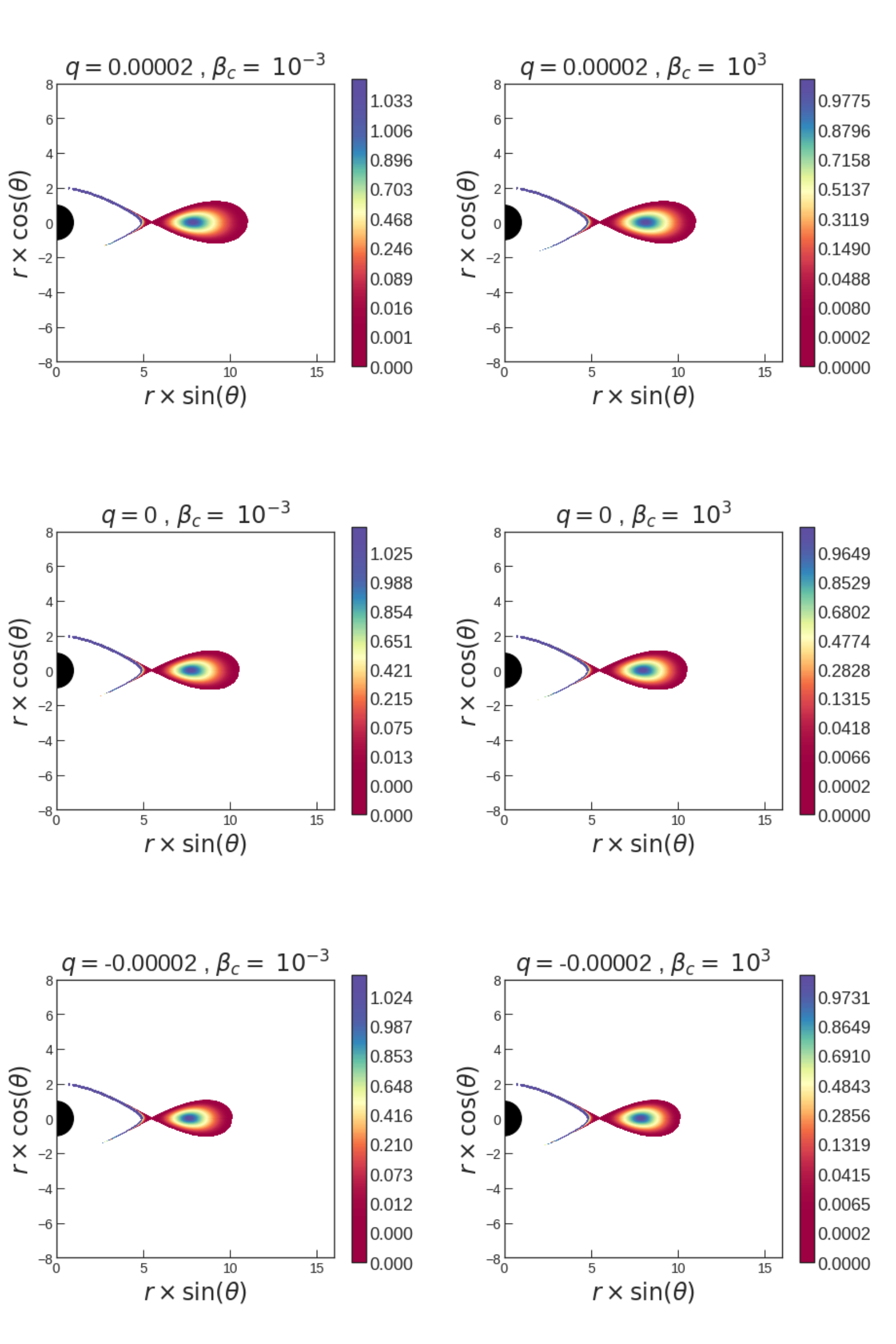}
    \caption{Map of the rest-mass density distribution for the more steeper angular momentum model $1$, as described in the Figure \ref{fig:6}. Each row corresponds to different values of $\rm q$, and each column shows different value of $\betac$.}
    \label{fig:10} 
  \end{figure}
  
In Figures \ref{fig:9} and \ref{fig:10} the rest-mass density distribution profiles are explored for a different combination of parameters to clearly see the effect of changing them. Figure \ref{fig:9} presents the results for the angular momentum distribution model $2$, and for the cases $G_2, H_2$, and $J_2$ in Tables \ref{tab:2} and \ref{tab:3}. In Figure \ref{fig:9} each column represents different magnetization parameters while changing quadrupole value exploring in rows. As we expected from the constant angular momentum case in the previous subsection \ref{constant}, magnetic pressure causes the rest of mass density to increase in the inner part of the disk regardless of the type distribution of angular momentum, as also seen in Figure \ref{fig:10}. Furthermore, the overall shape of the disk depends on the choice of $\rm q$. Especially the disk is more radially extended for positive values of $\rm q$. Also, further analysis shows for positive values of $\rm q$; the disk has more balloon shape than negative values. 

Figure \ref{fig:10} presents the result for for the angular momentum distribution model $1$, and for the cases $G_1, H_1$, and $J_1$ in Tables \ref{tab:2} and \ref{tab:3}. In Figure \ref{fig:10} we have the same patterns as in Figure \ref{fig:9}. However, for the distribution of model $1$, we obtain a smaller disk structure in both vertical and radial directions. Since as we have seen in Figure \ref{fig:6} the place of $x_{\rm c}$ and $x_{\rm cusp}$ are closer to each other. Also, comparing these Figures shows the smallest disk is obtained when the steepest distribution is combined with the largest quadrupole moment. Besides, we can see in each case, as the slope of specific angular momentum becomes steeper, the matter accumulates more in the inner part. Thus steeper the slope, the more centrally the matter distributes. This pattern also reminds the role of the magnetization parameter.

\subsection{discussion on Trigonometric distribution}

\begin{figure}[!htbp]
   \centering
\begin{tabular}{c}
\includegraphics[width=7.cm]{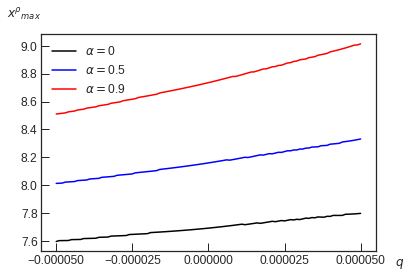}\\
\includegraphics[width=7.cm]{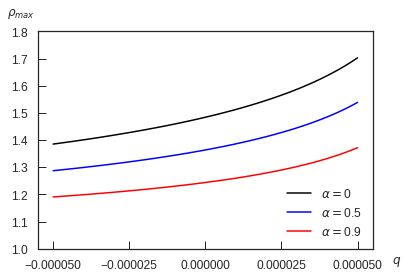}
\end{tabular}
    \caption{Amplitude and location of the rest-mass density maximum as a function of $\rm q$. The different curves represent different distributions of specific angular momentum.}
    \label{fig:11} 
\end{figure}

\begin{figure}[!htbp]
   \centering
   \begin{tabular}{c}
\includegraphics[width=7.cm]{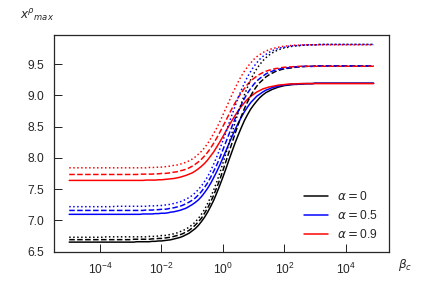}\\
\includegraphics[width=7.cm]{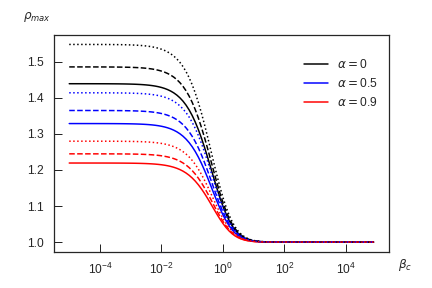}
\end{tabular}
    \caption{Amplitude and location of the rest-mass density maximum as a function of $\rm \betac$, and for different angular momentum profiles. The solid lines correspond to $\rm q= -0.00002$. The Schwarzschild case is shown in dashed line. Also, $\rm q= 0.00002$ is represented in dotted line.}
    \label{fig:12} 
\end{figure}

In this section, we discuss different profiles of $\ell(x,y)$ and values of parameter $\rm q$ following the method developed in \cite{2009A&A...498..471Q} which is described in Subsection \ref{5q}.
As in the previous case, we are interested in exploring the effects of the magnetization parameter $\beta_{\rm c}$, $\rm q$ and different profiles of angular momentum on the variation of the location and amplitude of the maximum rest-mass density. In Figures \ref{fig:11} and \ref{fig:12}, we have discussed the rest-mass density maximum and its place for different angular momentum profiles, $\rm q$ and $\beta_{\rm c}$. Like previous cases, increasing $\rm q$ causes an increase in rest-mass density maximum and its location. However, as $\alpha$ becomes higher for a fixed value of $\rm q$ and the curve deviates from the constant angular momentum distribution, the intensity in density maximum reduces, and the matter spreads more through the disk. Also, Figure \ref{fig:12} shows the sensitivity of each model on $\beta_{\rm c}$. Also, in this case, $\beta_{\rm c}$ has a stronger effect on the disk which its angular momentum distribution is closer to the constant one. And as the matter becomes more distributed through the disk caused by a higher value of $\alpha$, magnetic pressure also has a weaker effect on the disk.

\begin{figure*}[!htbp]
   \centering
\includegraphics[width=10cm]{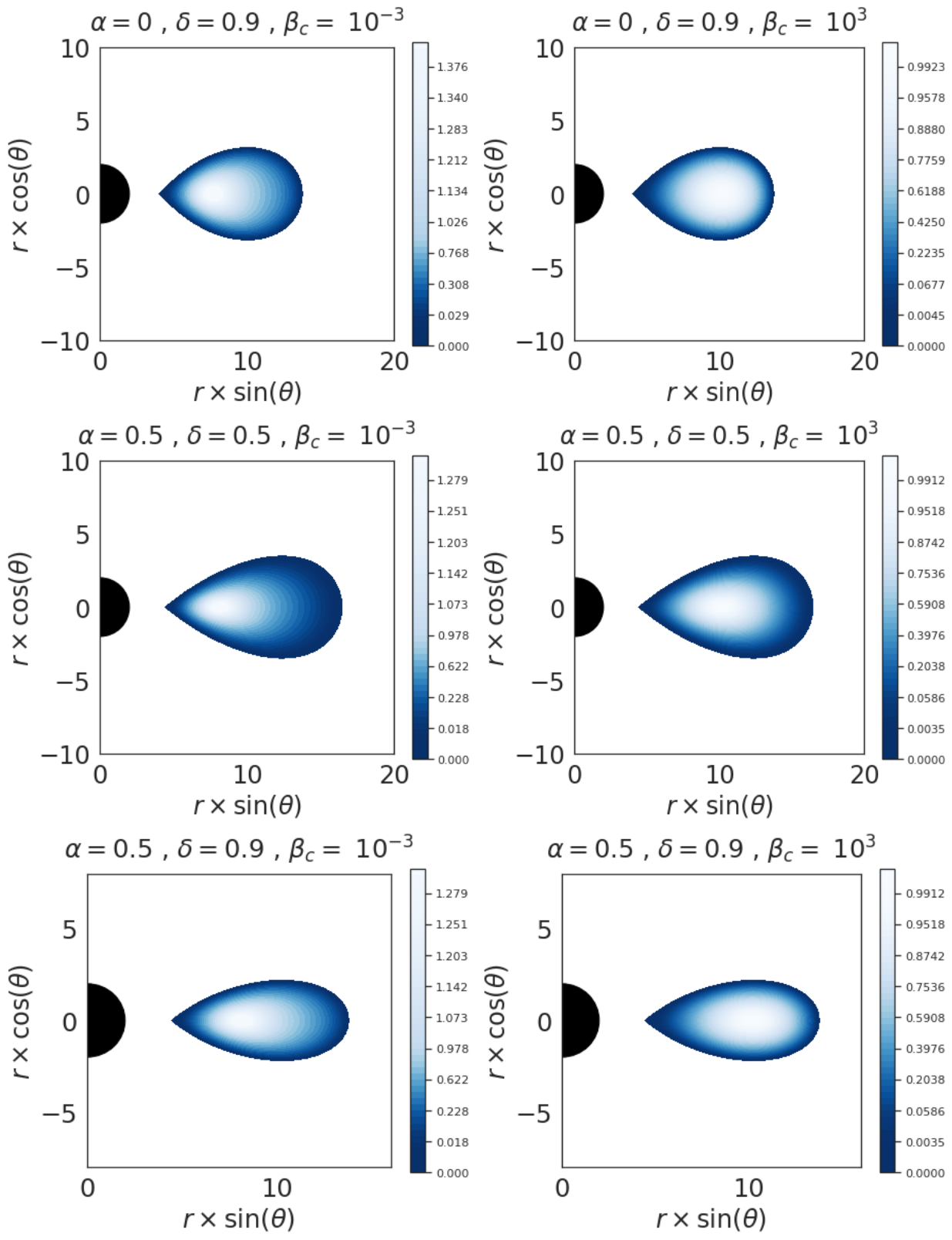}
    \caption{Effect of the magnetized parameter $\rm q$ on the rest-mass density distribution. The deformation parameter is fixed to be $\rm q=-0.00002$.}
    \label{fig:13} 
  \end{figure*}

\begin{figure*}[!htbp]
   \centering
\includegraphics[width=13cm]{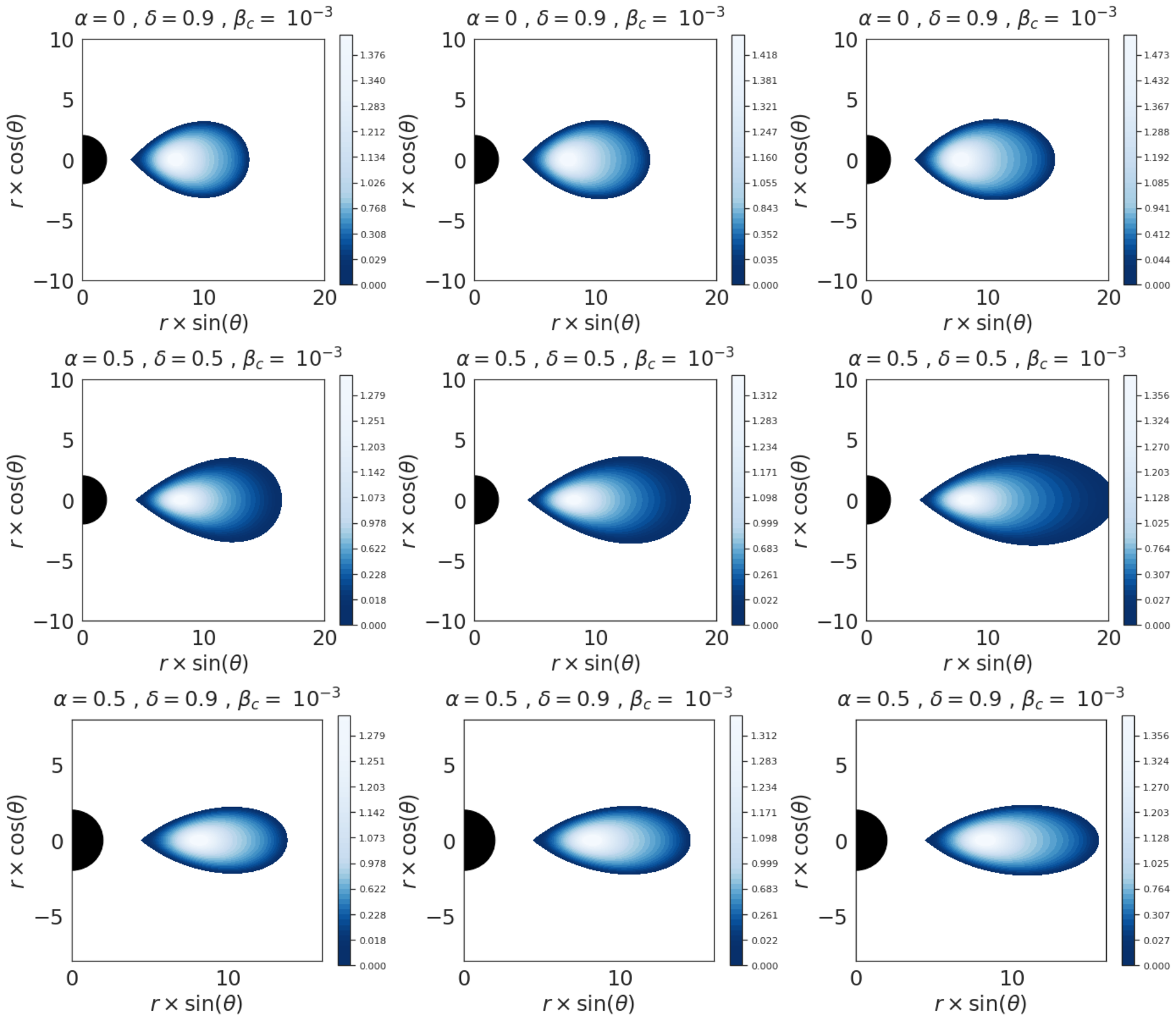}
    \caption{Effect of the distortion parameter $\rm q$ on the rest-mass density distribution. The first column is dedicated to $\rm q=-0.00002$. The middle column represents the Schwarzschild case $\rm q=0$. The last column shows $\rm q=0.00002$. The magnetization parameter is fixed to be $\betac=10^{-3}$.}
    \label{fig:14} 
  \end{figure*}

In Figure \ref{fig:13} and \ref{fig:14} the panels of rest-mess density distributions presented. The parameters are listed in Tables \ref{tab:Q1} and \ref{tab:Q2}. In Figure \ref{fig:13} the goal is to study the effect of parameter $\beta_{\rm c}$, and in the first column, we have relatively strong magnetic pressure compare to the second one. However, the model parameters are the same to distinguish the effect of magnetic pressure clearly. Also, $\rm q$ is set to be fixed. As mentioned before, $\beta_{\rm c}$ influences the distribution of mater inside the disk and not on the shape of the disk, in general. However, parameters $\alpha$ and $\delta$ have influenced the disk structure strongly. In fact, these two parameters are highly correlated to each other. However, further analysis shows that for a fixed $\alpha$, an increase in $\sigma$ causes the disk becomes thinner, the same for a fixed value of $\delta$.

Figure \ref{fig:14} explores the effects of different parameters, particularly $\rm q$, on the panel of rest-mass density in the presence of relatively strong magnetic pressure. Comparison of columns shows the effect of different values of $\rm q$ on the models. The parameter $\rm q$ is responsible for the radial extension of the disk as it becomes higher. In addition, a deeper analysis shows $\rm q$ is also responsible for a shift in the radial direction of the disk. In the sense that as the values of $\rm q$ increase, the disc shifts in the outward direction. However, as the values of quadrupole parameters that we chose in our models are minimal, this is not easy to see from the panels. This is also coherent with the shift in the place of the rest-mass density maximum as $\rm q$ increases.

\begin{table}
    \centering
\begin{tabular}{p{0.7cm}p{0.7cm}p{1.7cm}p{0.7cm}p{0.9cm}p{.9cm}p{.9cm}}
\hline
 $\alpha$ & $\delta$  & $\rm q$ & $\xc$ & $\xmaxP$ & $\xmaxPM$   \\
    \hline
    \hline
    $0$   & $0.9$   & $0.00002$ &  $9.191$ & $6.663$ & $7.230$  \\
 \hline
  $0.5$   & $0.5$   & $0.00002$ & $9.191$ & $7.077$ & $7.660$  \\
  \hline
  $0.5$   &$0.9$   & $0.00002$ & $9.191$ & $7.078$ & $7.662$  \\
  \hline
  \hline
  $0$   & $0.9$   & $0$ &  $9.476$ & $6.692$ & $7.352$  \\
 \hline
  $0.5$   & $0.5$   & $0$ &  $9.476$ & $7.128$ & $7.746$  \\
  \hline
  $0.5$   &$0.9$   & $0$ &  $9.476$ & $7.130$ & $7.746$  \\
 \hline
 \hline
 $0$   & $0.9$   & $-0.00002$ &   $9.814$ & $6.751$ & $7.446$  \\
 \hline
  $0.5$   & $0.5$   & $-0.00002$ & $9.814$ & $7.210$ & $7.858$  \\
  \hline
  $0.5$   &$0.9$   & $-0.00002$ &  $9.814$ & $7.210$ & $7.858$  \\
 \hline
\end{tabular}
    \caption{Parameters of the considered models for non-constant trigonometric specific angular momentum. All the models have $\beta_c=10^{-3}$.}
    \label{tab:Q1}
\end{table}

\begin{table}
    \centering
\begin{tabular}{p{0.7cm}p{0.7cm}p{1.7cm}p{0.7cm}p{0.9cm}p{.9cm}p{.9cm}}
\hline
 $\alpha$ & $\delta$  & $\rm q$ & $\xc$ & $\xmaxP$ & $\xmaxPM$   \\
    \hline
    \hline
    $0$   & $0.9$   & $0.00002$ &  $9.191$ & $9.191$ & $10.017$  \\
 \hline
  $0.5$   & $0.5$   & $0.00002$ & $9.191$ & $9.190$ & $10.016$  \\
  \hline
  $0.5$   &$0.9$   & $0.00002$ & $9.191$ & $9.190$ & $9.864$  \\
  \hline
  \hline
  $0$   & $0.9$   & $0$ &  $9.476$ & $9.476$ & $10.346$  \\
 \hline
  $0.5$   & $0.5$   & $0$ &  $9.476$ & $9.475$ & $10.305$  \\
  \hline
  $0.5$   &$0.9$   & $0$ &  $9.476$ & $9.476$ & $10.209$  \\
  \hline
 \hline
 $0$   & $0.9$   & $-0.00002$ &   $9.814$ & $9.814$ & $10.722$  \\
 \hline
  $0.5$   & $0.5$   & $-0.00002$ & $9.814$ &  $9.814$ & $11.000$  \\
  \hline
  $0.5$   &$0.9$   & $-0.00002$ &  $9.814$ &  $9.814$ & $10.701$  \\
  \hline
 \hline
\end{tabular}
    \caption{The same cases as Table \ref{tab:Q1} for $\beta_c=10^{3}$.}
    \label{tab:Q2}
\end{table}

\section{Summary and conclusion}\label{sec:discuss}
In this paper, we analyzed equilibrium sequences of magnetized, non-self-gravitating disks around distorted Schwarzschild black holes. In fact, this is a solution to the Einstein equation, which is a vacuum and static with a regular horizon but not asymptotically flat and is obtained by assuming the existence of an external distribution of matter outside the horizon \cite{1982JMP....23..680G}. 

In this procedure, we combine the two existing approaches of \citep{Komissarov_2006} combined with \citep{2009A&A...498..471Q}, also \citep{Komissarov_2006} combined with \citep{2015MNRAS.447.3593W} to study the properties of the thick disk structure in this background. We also compared the results with the Schwarzschild case. Explicitly, on the one hand, we followed the method in \citep{Komissarov_2006} to attach a dynamically toroidal magnetic field to the model. On the other hand, besides specific constant angular momentum distribution, we consider two models for the non-constant distribution of angular momentum where the location and the morphology of the surfaces of equipotential can be computed, as introduced in \cite{2009A&A...498..471Q} and \cite{2015MNRAS.447.3593W}, which have been briefly explained in Section \ref{sec2}. As an especial case, the results are in good agreement with the result of \cite{Komissarov_2006} when we limited our solution to the constant angular momentum distribution with the vanishing quadrupole moment. 

We have analyzed these models' properties and their dependency on the magnetization parameter $\betac$, the quadrupole parameter $\rm q$, and the angular momentum distribution. We have shown the effects of changing parameter $\betac$, in general, does not change the disk's overall configuration in both constant and non-constant angular momentum distributions. However, the strongest effect of $\betac$ was seen on the rest-mass density in the disk's inner part. Furthermore, a slight change occurs in the range of isodensity contours, which is also compatible with increasing density toward the disk's inner part.

On the other hand, the effect of varying quadrupole moments has changed the disk's geometry and overall shape. In addition, this changes the location of maximum rest-mass density and its value. We traced the fingerprint of different angular momentum distribution, the shape and size of the disk, and the rest-mass density profile.


This work can be seen as the first step toward considering the self-gravity of the disk, which is the plan for extending this work. Besides, one can study different forms of the barotropic equation of state. This is also interesting to use these analytical models as the initial conditions as a seed for numerical simulation and test their ability to account for observable constraints of astrophysical systems.

\section {Acknowledgements}
The authors are grateful to the anonymous referee for the useful comments. This work is supported by the research training group GRK 1620 ”Models of Gravity”, funded by the German Research Foundation (DFG).

\bibliographystyle{unsrt}
\bibliography{nameyy}

\newcommand{\noop}[1]{}
\begin{thebibliography}{10}

\bibitem{1974AcA....24...45A}
M.~A. {Abramowicz}.
\newblock {Theory of Level Surfaces Inside Relativistic: Rotating Stars. II.}
\newblock {\em \actaa}, 24:45, January 1974.

\bibitem{1978A&A....63..209K}
M.~{Kozlowski}, M.~{Jaroszynski}, and M.~A. {Abramowicz}.
\newblock {The analytic theory of fluid disks orbiting the Kerr black hole.}
\newblock {\em \aap}, 63(1-2):209--220, February 1978.

\bibitem{1980AcA....30....1J}
M.~{Jaroszynski}, M.~A. {Abramowicz}, and B.~{Paczynski}.
\newblock {Supercritical accretion disks around black holes}.
\newblock {\em \actaa}, 30(1):1--34, January 1980.

\bibitem{1980A&A....88...23P}
B.~{Paczy{\'n}sky} and P.~J. {Wiita}.
\newblock {Thick accretion disks and supercritical luminosities.}
\newblock {\em \aap}, 500:203--211, August 1980.

\bibitem{1980ApJ...242..772A}
M.~A. {Abramowicz}, M.~{Calvani}, and L.~{Nobili}.
\newblock {Thick accretion disks with super-Eddington luminosities}.
\newblock {\em \apj}, 242:772--788, December 1980.

\bibitem{1981Natur.294..235A}
M.~A. {Abramowicz}.
\newblock {Innermost parts of accretion disks are thermally and secularly
  stable}.
\newblock {\em \nat}, 294(5838):235--236, November 1981.

\bibitem{1982MitAG..57...27P}
B.~{Paczynski}.
\newblock {Thick Accretion Disks around Black Holes
  (Karl-Schwarzschild-Vorlesung 1981)}.
\newblock {\em Mitteilungen der Astronomischen Gesellschaft Hamburg}, 57:27,
  January 1982.

\bibitem{1982ApJ...253..897P}
B.~{Paczynski} and M.~A. {Abramowicz}.
\newblock {A model of a thick disk with equatorial accretion}.
\newblock {\em \apj}, 253:897--907, February 1982.

\bibitem{1988ApJ...327..116M}
Piero {Madau}.
\newblock {Thick Accretion Disks around Black Holes and the UV/Soft X-Ray
  Excess in Quasars}.
\newblock {\em \apj}, 327:116, April 1988.

\bibitem{1996ApJ...458..474S}
Ewa {Szuszkiewicz}, Matthew~A. {Malkan}, and Marek~A. {Abramowicz}.
\newblock {The Observational Appearance of Slim Accretion Disks}.
\newblock {\em \apj}, 458:474, February 1996.

\bibitem{1984ApJ...277..296H}
J.~F. {Hawley}, L.~L. {Smarr}, and J.~R. {Wilson}.
\newblock {A numerical study of nonspherical black hole accretion. I Equations
  and test problems}.
\newblock {\em \apj}, 277:296--311, February 1984.

\bibitem{1984ApJS...55..211H}
J.~F. {Hawley}, L.~L. {Smarr}, and J.~R. {Wilson}.
\newblock {A numerical study of nonspherical black hole accretion. II - Finite
  differencing and code calibration}.
\newblock {\em \apjs}, 55:211--246, June 1984.

\bibitem{2001ApJ...554L..49H}
John~F. {Hawley}, Steven~A. {Balbus}, and James~M. {Stone}.
\newblock {A Magnetohydrodynamic Nonradiative Accretion Flow in Three
  Dimensions}.
\newblock {\em \apjl}, 554(1):L49--L52, June 2001.

\bibitem{2003ApJ...592.1060D}
Jean-Pierre {De Villiers} and John~F. {Hawley}.
\newblock {Global General Relativistic Magnetohydrodynamic Simulations of
  Accretion Tori}.
\newblock {\em \apj}, 592(2):1060--1077, August 2003.

\bibitem{1991ApJ...376..214B}
Steven~A. {Balbus} and John~F. {Hawley}.
\newblock {A Powerful Local Shear Instability in Weakly Magnetized Disks. I.
  Linear Analysis}.
\newblock {\em \apj}, 376:214, July 1991.

\bibitem{Komissarov_2006}
S.~S. Komissarov.
\newblock Magnetized tori around kerr black holes: analytic solutions with a
  toroidal magnetic field.
\newblock {\em Monthly Notices of the Royal Astronomical Society},
  368(3):993–1000, Apr 2006.

\bibitem{2015A&A...574A..48V}
F.~H. {Vincent}, W.~{Yan}, O.~{Straub}, A.~A. {Zdziarski}, and M.~A.
  {Abramowicz}.
\newblock {A magnetized torus for modeling Sagittarius A$^{{\ensuremath{*}}}$
  millimeter images and spectra}.
\newblock {\em \aap}, 574:A48, January 2015.

\bibitem{2007MNRAS.378.1101M}
P.~J. {Montero}, O.~{Zanotti}, J.~A. {Font}, and L.~{Rezzolla}.
\newblock {Dynamics of magnetized relativistic tori oscillating around black
  holes}.
\newblock {\em \mnras}, 378(3):1101--1110, July 2007.

\bibitem{2009A&A...498..471Q}
Lei {Qian}, M.~A. {Abramowicz}, P.~C. {Fragile}, J.~{Hor{\'a}k}, M.~{Machida},
  and O.~{Straub}.
\newblock {The Polish doughnuts revisited. I. The angular momentum distribution
  and equipressure surfaces}.
\newblock {\em \aap}, 498(2):471--477, May 2009.

\bibitem{2015MNRAS.447.3593W}
Maciek {Wielgus}, P.~Chris {Fragile}, Ziming {Wang}, and Julia {Wilson}.
\newblock {Local stability of strongly magnetized black hole tori}.
\newblock {\em \mnras}, 447(4):3593--3601, March 2015.

\bibitem{2017MNRAS.467.1838F}
P.~Chris {Fragile} and Aleksander {Sadowski}.
\newblock {On the decay of strong magnetization in global disc simulations with
  toroidal fields}.
\newblock {\em \mnras}, 467(2):1838--1843, May 2017.

\bibitem{2017A&A...607A..68G}
Sergio {Gimeno-Soler} and Jos{\'e}~A. {Font}.
\newblock {Magnetised Polish doughnuts revisited}.
\newblock {\em \aap}, 607:A68, November 2017.

\bibitem{Kovar11}
J.~{Kov{\'a}{\v r}}, P.~{Slan{\'y}}, Z.~{Stuchl{\'{\i}}k}, V.~{Karas},
  C.~{Cremaschini}, and J.~C. {Miller}.
\newblock {Role of electric charge in shaping equilibrium configurations of
  fluid tori encircling black holes}.
\newblock {\em \prd}, 84(8):084002, oct 2011.

\bibitem{KoSlaCreStuKaTro16}
J.~{Kov{\'a}{\v r}}, P.~{Slan{\'y}}, C.~{Cremaschini}, Z.~{Stuchl{\'{\i}}k},
  V.~{Karas}, and A.~{Trova}.
\newblock {Charged perfect fluid tori in strong central gravitational and
  dipolar magnetic fields}.
\newblock {\em \prd}, 93(12):124055, jun 2016.

\bibitem{Stuchlik2004}
Z.~{Stuchl{\'{\i}}k} and P.~{Slan{\'y}}.
\newblock {Accretion disks in the Kerr-de Sitter spacetimes}.
\newblock In S.~{Hled{\'{\i}}k} and Z.~{Stuchl{\'{\i}}k}, editors, {\em RAGtime
  4/5: Workshops on black holes and neutron stars}, pages 205--237, December
  2004.

\bibitem{Kovacs2010}
Z.~{Kov{\'a}cs} and T.~{Harko}.
\newblock {Can accretion disk properties observationally distinguish black
  holes from naked singularities?}
\newblock {\em \prd}, 82(12):124047, Dec 2010.

\bibitem{Harko2009}
Tiberiu {Harko}, Zolt{\'a}n {Kov{\'a}cs}, and Francisco S.~N. {Lobo}.
\newblock {Thin accretion disks in stationary axisymmetric wormhole
  spacetimes}.
\newblock {\em \prd}, 79(6):064001, Mar 2009.

\bibitem{2003A&A...412..603R}
L.~{Rezzolla}, O.~{Zanotti}, and J.~A. {Font}.
\newblock {Dynamics of thick discs around Schwarzschild-de Sitter black holes}.
\newblock {\em \aap}, 412:603--613, December 2003.

\bibitem{2005CQGra..22.3623S}
Petr {Slan{\'y}} and Zdenek {Stuchl{\'\i}k}.
\newblock {Relativistic thick discs in the Kerr de Sitter backgrounds}.
\newblock {\em Classical and Quantum Gravity}, 22(17):3623--3651, September
  2005.

\bibitem{2011JCAP...01..033K}
Hana {Kuc{\'a}kov{\'a}}, Petr {Slan{\'y}}, and Zden{\u{e}}k {Stuchl{\'\i}k}.
\newblock {Toroidal configurations of perfect fluid in the
  Reissner-Nordstr{\"o}m-(anti-)de Sitter spacetimes}.
\newblock {\em \jcap}, 2011(1):033, January 2011.

\bibitem{2021JCAP...03..063T}
Matheus~C. {Teodoro}, Lucas~G. {Collodel}, and Jutta {Kunz}.
\newblock {Retrograde polish doughnuts around boson stars}.
\newblock {\em \jcap}, 2021(3):063, March 2021.

\bibitem{1982JMP....23..680G}
R.~{Geroch} and J.~B. {Hartle}.
\newblock {Distorted black holes.}
\newblock {\em Journal of Mathematical Physics}, 23:680--692, 1982.

\bibitem{Chandrasekhar:579245}
S~Chandrasekhar.
\newblock {\em {The mathematical theory of black holes}}.
\newblock Oxford classic texts in the physical sciences. Oxford Univ. Press,
  Oxford, 2002.

\bibitem{1965ZhETF...49.170D}
A.~G. {Doroshkevich}, Ya.~B. {Zel'dovich}, and I.~D. {Novikov}.
\newblock {Gravitational Collapse of Non-Symmetric and Rotating Bodies}.
\newblock {\em Zhurnal Eksperimentalnoi i Teoreticheskoi Fiziki}, 49:170,
  December 1965.

\bibitem{2020PhRvD.101b3002F}
Shokoufe {Faraji} and Eva {Hackmann}.
\newblock {Thin accretion disk around the distorted Schwarzschild black hole}.
\newblock {\em \prd}, 101(2):023002, January 2020.

\bibitem{PhysRevD.78.024040}
Jos\'e P.~S. Lemos and Oleg~B. Zaslavskii.
\newblock Black hole mimickers: Regular versus singular behavior.
\newblock {\em Phys. Rev. D}, 78:024040, Jul 2008.

\bibitem{2019MNRAS.482...52S}
Rajibul {Shaikh}, Prashant {Kocherlakota}, Ramesh {Narayan}, and Pankaj~S.
  {Joshi}.
\newblock {Shadows of spherically symmetric black holes and naked
  singularities}.
\newblock {\em \mnras}, 482(1):52--64, January 2019.

\bibitem{2002A&A...396L..31A}
M.~A. {Abramowicz}, W.~{Klu{\'z}niak}, and J.~P. {Lasota}.
\newblock {No observational proof of the black-hole event-horizon}.
\newblock {\em \aap}, 396:L31--L34, December 2002.

\bibitem{1976JMP....17...54E}
F.~J. {Ernst}.
\newblock {Black holes in a magnetic universe}.
\newblock {\em Journal of Mathematical Physics}, 17(1):54--56, January 1976.

\bibitem{1976JMP....17..182E}
Frederick~J. {Ernst} and Walter~J. {Wild}.
\newblock {Kerr black holes in a magnetic universe}.
\newblock {\em Journal of Mathematical Physics}, 17(2):182--184, February 1976.

\bibitem{PhysRevD.10.1680}
Robert~M. Wald.
\newblock Black hole in a uniform magnetic field.
\newblock {\em Phys. Rev. D}, 10:1680--1685, Sep 1974.

\bibitem{Note1}
Boyer's condition states the boundary of any stationary and barotropic perfect
  fluid body is an equipotential surface.

\bibitem{1978A&A....63..221A}
M.~{Abramowicz}, M.~{Jaroszynski}, and M.~{Sikora}.
\newblock {Relativistic, accreting disks.}
\newblock {\em \aap}, 63:221--224, February 1978.

\bibitem{1924MNRAS..84..665V}
H.~{von Zeipel}.
\newblock {The radiative equilibrium of a rotating system of gaseous masses}.
\newblock {\em \mnras}, 84:665--683, June 1924.

\bibitem{2015GReGr..47...44Z}
O.~{Zanotti} and D.~{Pugliese}.
\newblock {Von Zeipel's theorem for a magnetized circular flow around a compact
  object}.
\newblock {\em General Relativity and Gravitation}, 47:44, April 2015.

\bibitem{1978srfm.book.....D}
W.~G. {Dixon}.
\newblock {\em {Special relativity: the foundation of macroscopic physics.}}
\newblock Oxford Univ. Press, 1978.

\bibitem{1989rfmw.book.....A}
Angelo~Marcello {Anile}.
\newblock {\em {Relativistic fluids and magneto-fluids : with applications in
  astrophysics and plasma physics}}.
\newblock Oxford Univ. Press, 1989.

\bibitem{1989PASJ...41..133O}
Rika {Okada}, Jun {Fukue}, and Ryoji {Matsumoto}.
\newblock {A model of astrophysical tori with magnetic fields}.
\newblock {\em \pasj}, 41(1):133--140, January 1989.

\bibitem{1997PhLA..230....7B}
Nora {Bret{\'o}n}, Tatiana~E. {Denisova}, and Vladimir~S. {Manko}.
\newblock {A Kerr black hole in the external gravitational field}.
\newblock {\em Physics Letters A}, 230:7--11, Feb 1997.

\bibitem{Note2}
The morphology of the solutions for nondistorted case indicates that there is
  no significant change in the solutions for magnetization values above $\beta
  _{\protect \rm c}=10^{3}$ and below $\beta _{\protect \rm c}=10^{-3}$.

\end{thebibliography}

\end{document}